\documentclass[aps,pra,showpacs,amsmath,amssymb,twocolumn,superscriptaddress]{revtex4-1}

\usepackage{graphicx}
\usepackage{dcolumn}
\usepackage{bm}
\usepackage{enumerate}
\usepackage{amsmath}
\usepackage{amssymb}
\usepackage{hyperref}
\usepackage{color}
\usepackage{times}
\usepackage{physics}
\usepackage{placeins}

\allowbreak

\begin{document}

\title{Fast and robust quantum state transfer via a topological chain}

\author{N. E. Palaiodimopoulos}
\email[]{nikpalaio@phys.uoa.gr}
\affiliation{Department of Physics, National and Kapodistrian University of Athens, GR-15784 Athens, Greece}

\author{I. Brouzos}
\affiliation{LAUM, UMR-CNRS 6613, Le Mans Universit\'e, Av. O. Messiaen, 72085, Le Mans, France}

\author{F. K. Diakonos}
\affiliation{Department of Physics, National and Kapodistrian University of Athens, GR-15784 Athens, Greece}

\author{G. Theocharis}
\affiliation{LAUM, UMR-CNRS 6613, Le Mans Universit\'e, Av. O. Messiaen, 72085, Le Mans, France}

\begin{abstract}
We propose a fast and robust quantum state transfer protocol employing a Su-Schrieffer-Heeger chain, where the interchain couplings vary in time. Based on simple considerations around the terms involved in the definition of the adiabatic invariant, we construct an exponential time-driving function that successfully takes advantage of resonant effects to speed up the transfer process. Using optimal control theory, we confirm that the proposed time-driving function is close to optimal. To unravel the crucial aspects of our construction, we proceed to a comparison with two other protocols. One where the underlying Su-Schrieffer-Heeger chain is adiabatically time-driven and another where the underlying chain is topologically trivial and resonant effects are at work. By numerically investigating the resilience of each protocol to static noise, we highlight the robustness of the exponential driving. 
\end{abstract}

\maketitle

\section{Introduction}

Constructing a quantum network where states can be transferred in a coherent manner between two nodes is a task of fundamental importance towards the realization of an efficient platform for quantum information processing \cite{divincenzo2000physical}. The past two decades great effort has been made to obtain the optimal protocol for state transfer in the simplest geometry, the one-dimensional quantum chain. The most common model describing a quantum chain is the one-dimensional spin-$1/2$ chain. The Hamiltonian used is quite generic and the results can be applied to a variety of physical systems such as evanescently coupled waveguides \cite{bellec2012faithful,perez2013coherent,chapman2016experimental}, acoustic cavities \cite{shen2019one}, diamond vacancies \cite{yao2011robust}, superconducting circuits \cite{tsomokos2010using,mei2018robust}, arrays of quantum dots \cite{petrosyan2006coherent}, driven optical lattices \cite{chen2011controlling}, NMR \cite{cappellaro2007simulations} and nanoelectromechanical networks \cite{tian2020perfect}. 

Depending on whether the parameters of the system vary in time or not the quantum state transfer (QST) protocols can be divided into two classes, time-independent and time-dependent. In the former, the parameters are initially engineered in a suitable manner and as the system evolves "freely" the transfer takes place. The protocols in this approach usually rely either on the seminal work initiated by Bose \cite{bose2003quantum} and later evolved in \cite{christandl2004perfect,nikolopoulos2004coherent}, or on works where the states are transferred via Rabi-like oscillation schemes \cite{wojcik2005unmodulated,oh2011heisenberg,giorgi2013quantum}. On the contrary, in time-dependent protocols, the system's parameters are controlled during the dynamical evolution. The most intuitive protocol in this case is to apply a sequence of swap operations between adjacent sites and gradually move the state along the chain, while other representative protocols are introduced in\cite{balachandran2008adiabatic,schirmer2009fast,burgarth2010scalable,korzekwa2014quantum}.

Besides the feasibility of the experimental implementation and the scalability, there are two major factors that determine the efficiency of a QST protocol. Namely, how much time it takes for the transfer to occur and how faithfully the state is transferred in the presence or absence of decoherence and static imperfections. The quantum speed limit for transferring a state along a spin chain has been studied for various protocols \cite{deffner2017quantum,yung2006quantum,ashhab2012speed,caneva2009optimal,zhang2018automatic} . On the other hand, many works \cite{de2005perfect,kay2006perfect,petrosyan2010state,bruderer2012exploiting} have examined the role of different sources of decoherence in QST protocols and proposed schemes \cite{burgarth2005conclusive,balachandran2008adiabatic,huang2018quantum,agundez2017superadiabatic,allcock2009quantum} to circumvent their impact. In most cases there is a trade-off between speed and robustness, as increasing one results to the decrease of the other and vice versa. 

A very promising direction towards the realization of an efficient platform able to perform fault-tolerant quantum computation comes from the flourishing field of topological states of matter \cite{sarma2006topological}. One of the most appealing properties of topological systems is that they host edge states which, due to their topological protection, are robust to different sources of quantum decoherence. Recent studies \cite{lang2017topological,estarellas2017topologically,mei2018robust,boross2019poor,longhi2019topological,longhi2019landau} have employed 1-D topological systems, such as the Kitaev chain \cite{kitaev2001unpaired} and the SSH model \cite{asboth2016short}, to act as a platform for realizing QST protocols. In this work, in the same spirit and aiming to balance the trade-off between the various factors that determine the efficiency of QST protocols, we propose a fast and robust protocol for transferring an excitation along an SSH chain. We consider a protocol where the exchange interaction between adjacent sites can vary with time. In order to reveal the crucial characteristics that favor our proposal we compare it with two other QST protocols, one \cite{mei2018robust} that employs a topologically non-trivial chain and one employing a topologically trivial chain. In the former the underlying chain is adiabatically driven, while in the later resonant processes are at work. 

The rest of the paper organizes as follows. In Sec. \ref{Protocols} we present the Hamiltonian of the system together with the corresponding protocols for both topological and topologically trivial quantum channels. In Sec. \ref{crucial} we identify the crucial characteristics the driving function needs to posses, in order to speed up QST in a topological quantum channel. In Sec. \ref{speed} we present numerical evidence supporting our claims. In Sec. \ref{Disorder} we analyze the impact of on and off-diagonal static noise. Finally, in Sec. \ref{Conclusions} we conclude.

\section{QST Protocols} \label{Protocols} 
We start by considering a paradigm model describing a spin-$1/2$ chain acting as a data bus for transferring a quantum state. The Hamiltonian describes $N$ spins, where nearest neighbors are coupled with an XX Heisenberg exchange interaction of strength $J_{i}$ and a local magnetic field $B_{i}$ is applied at each spin. When we restrict ourselves to the one-excitation subspace, where all spins point down but one, the Hamiltonian writes as follows: 
\begin{equation} \label{eq:1}
\mathcal{H}=\sum_{i=1}^{N-1} J_{i} ( \ket{i} \bra{i+1} + h.c.) + \sum_{i=1}^{N} B_{i} \ket{i} \bra{i}
\end{equation}
where $J_{i}, B_{i}\in \mathbb{R}$, $J_{i}\geq 0$ and $\ket{i}$ denotes that the $i$-th site of the chain is excited (i.e. in Fock space notation $\ket{0_{1} 0_{2} ... 1_{i} ...0_{N} }$). The aim of the protocols we will consider is to transfer a single site excitation from the first $\ket{1}$ to the last $\ket{N}$ site of the chain by properly controlling the couplings $J_{i}$ during the dynamical evolution. The quantity that determines how faithfully the transfer has occurred is the fidelity, which in our case can be defined as:
\begin{equation} \label{eq:2}
\mathcal{F}=\abs{\bra{N}\ket{N(t^{*})}}^2
\end{equation} 
where by $\ket{N(t^{*})}$ we denote amplitude of the $N$-th site, obtained by numerically solving the time-dependent Schrodinger equation and $t^{*}$ corresponds to the transfer time. We must note here, that the fidelity for transferring a generic initial state of the form $\psi_{init}=\cos(\theta/2) \ket{0}+e^{i\phi} \sin(\theta/2) \ket{1}$ from the first to the last site of the chain, is given by $F=\frac{1}{3}+\frac{1}{6}(1+\mathcal{F})^{2}$ \cite{bose2003quantum}. However, since $F$ is simply a function of $\mathcal{F}$, throughout this paper, unless explicitly stated otherwise, whenever we refer to fidelity we will consider the quantity of Eq. \ref{eq:2}. 

In all protocols we will present, the energy scale is determined by the maximum value that the couplings acquire during the dynamical evolution. Thus, without loss of generality we will set $J_{max}=1$. Time will be given in units of $1/J_{max}$ and energy in units of $J_{max}$. We will now present two different cases for realizing an efficient time-driven quantum channel, one where the underlying undriven chain has topological characteristics and another where the chain is topologically trivial.

\subsection{Topological chain}

\FloatBarrier
The SSH model is the simplest topologically non-trivial system in 1-D that can be obtained by suitably modifying the Hamiltonian of Eq. \ref{eq:1}. To do so, the chain has to be dimerized. Meaning, we have to make $J_{i}=J_{odd}$ for $i \in odd$ and $J_{i}=J_{even}$ for $i \in even$. In addition, the magnetic field needs to be constant. We will assume that $B_{i}=0$ $\forall i$. For even-sized chains the topological phase arises when $J_{odd} > J_{even}$ and two edge modes appear at the two ends of the chain. The energies corresponding to the two eigenmodes lie close (above and below) to $E=0$ and are separated from the rest of the modes by a finite energy gap. The size of the gap is related to the ratio between $J_{even}$ and $J_{odd}$ \cite{asboth2016short}. On the contrary, for odd-sized chains, there is always one edge mode that is localized on the end corresponding to the weaker coupling. Namely, when $J_{odd} < J_{even}$ ($J_{odd} >J_{even}$) the mode is localized near the first (last) site of the chain. In this case the energy of the mode is exactly zero. For odd-sized chains the eigenmode energies (besides the zero energy solution) are given by the following expression: 
\begin{equation}
\label{eq:pairs}
\epsilon_{j}=\pm \abs{J_{odd}+J_{even} e^{i q_{j}}},
\end{equation} 
where $q_{j}=2 j \pi / (N+1)$ and $j=1, 2,..., [N/2]$ ($j$ counts the number of $\pm$ pairs and $[x]$ gives the greatest integer that is less than equal to $x$) \cite{coutant2020robustness}. Thus, for an odd-sized chain of length $N$ the energy gap can be analytically determined and is given by:
\begin{equation}
\label{eq:en}
g=2 \abs{\epsilon_{[N/2]}}
\end{equation}

\begin{figure}
\center
\includegraphics[width=0.4 \textwidth]{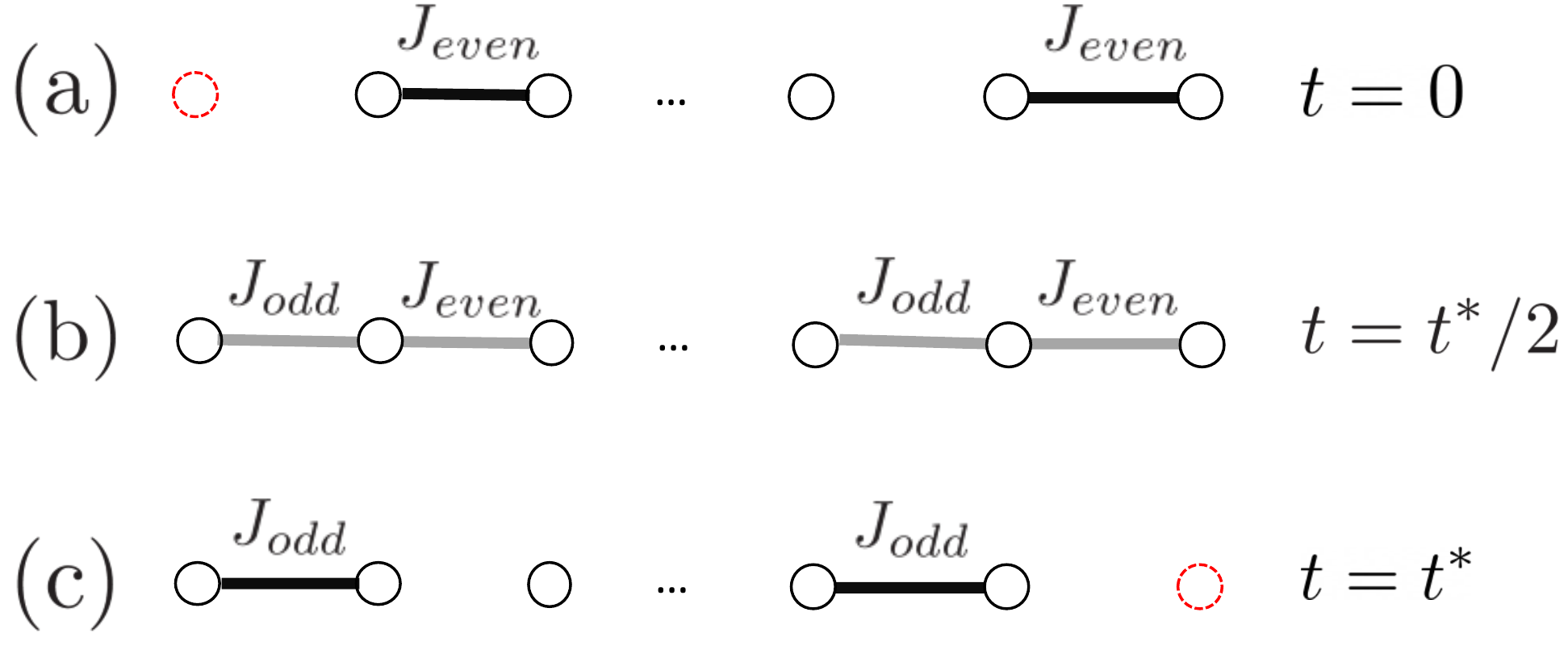}\caption{\label{fig:1a} A schematic of different time instants during the dynamical evolution of the topological chain. (a) Initially, $J_{odd}=0$, $J_{even}=J_{max}$, the zero-mode is localized on the first site and the energy gap takes its maximum value. (b) For $t=t^{*}/2$ we have $J_{1}=J_{N-1}<J_{max}$ and the energy gap acquires its minimum value. Before and after $t^{*}/2$ we have $J_{odd}<J_{even}<J_{max}$ and $J_{even}<J_{odd}<J_{max}$ respectively. (c) Finally, $J_{odd}=J_{max}$, $J_{even}=0$, the zero-mode is localized on the last site and the energy gap, once again takes its maximum value.}
\centering
\end{figure}

\FloatBarrier

For the protocols we will consider in this case, we restrict ourselves to odd-sized chains and we assume that we can separately control even and odd-indexed couplings. Initially the system is prepared so that $J_{odd}=0$ and $J_{even}=1$ and the excitation is localized on the first site of the chain which is disconnected from the rest (see Fig. \ref{fig:1a} (a)). Therefore, the initial state is an eigenstate of the system ($\ket{1 0 0 ... 0}$ wth zero eigenenergy). At the transfer time $t^{*}$ (see Fig. \ref{fig:1a} (c)) we end up with the reverse situation (i.e. $J_{even}=0$ and $J_{odd}=1$). The system undergoes a transition and transforms from a topological chain supporting an edge mode on the first site, to a topological chain with an edge mode on the last site, resulting to an excitation transfer from one side to the other. Note here, that in this protocol, there always exists a time where all the couplings acquire the same value $J_{even}=J_{odd}$. In our case we will consider this time to be $t=t^{*}/2$ (see Fig. \ref{fig:1a} (b)). For the infinite system, $J_{even}=J_{odd}$ corresponds to the closing of the energy gap separating the zero-energy mode with the rest excited states. However, in finite systems a finite energy difference between any two modes is always present. Thus, the point in the parameter space where $J_{even}=J_{odd}$, corresponds to the minimization of the energy gap.

\subsection{Topologically-trivial chain}
\begin{figure}
\center
\includegraphics[width=0.4 \textwidth]{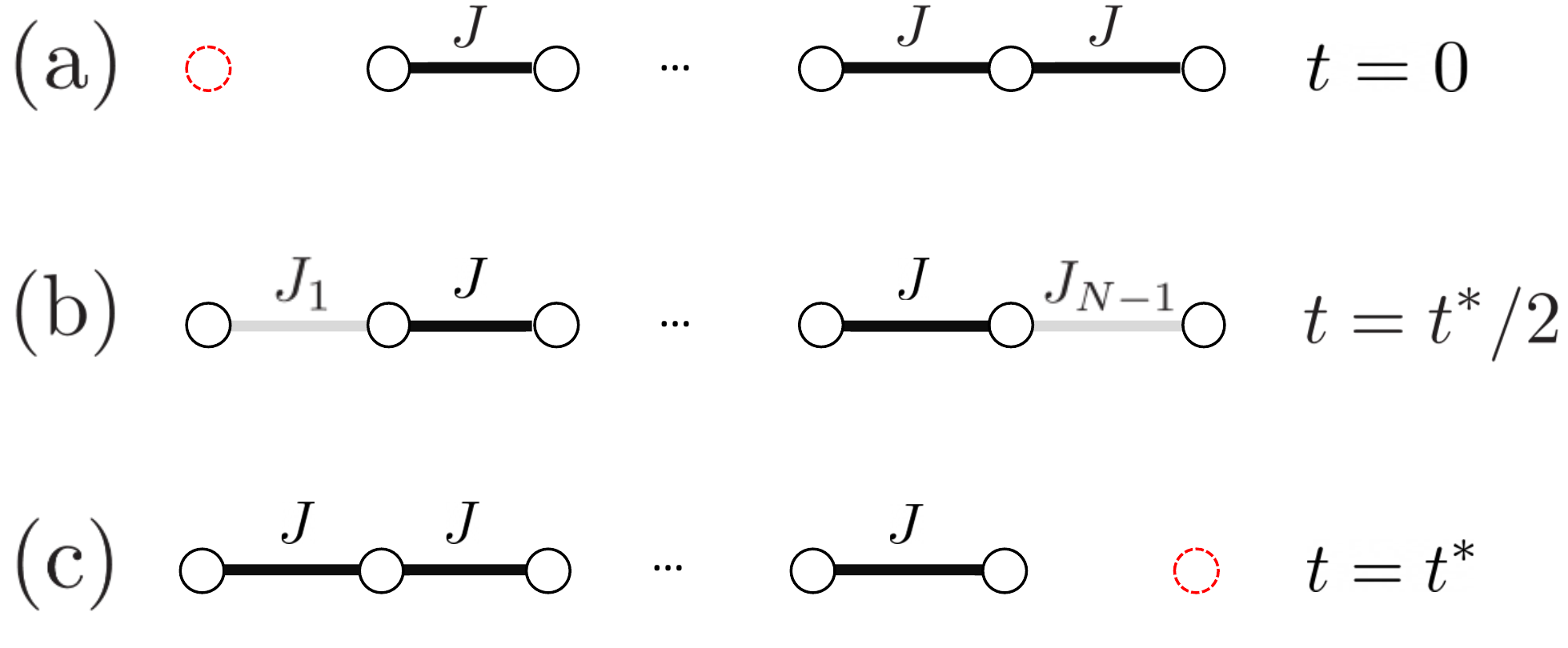}\caption{\label{fig:1b} A schematic of different time instants during the dynamical evolution of the topologically-trivial chain. (a) Initially, $J_{1}=0$, $J_{i}=J_{max}$ where $i=2,...,N-1$, the zero-mode is localized on the first site and the energy gap takes its maximum value. (b) For $t=t^{*}/2$ we have $J_{1}=J_{N-1}<J_{max}$ while $J_{i}=J_{max}$ for $i=2,3,...,J_{N-2}$. Before and after $t^{*}/2$ we have $J_{1}<J_{N-1}<J_{max}$ and $J_{N-1}<J_{1}<J_{max}$ respectively. (c) Finally, $J_{1}=0$ and $J_{i}=J_{max}$ for $i=2,...,N-1$, while the zero-mode is localized at the last site of the chain.}
\centering
\end{figure}

To explore how the topological nature of the underlying static chain affects the transfer we also proceed to a comparison with a protocol employing a topologically-trivial chain. We consider a protocol where the only couplings that are controlled are the ones connecting the edge sites with the rest of the chain (i.e. $J_{1}$ and $J_{N-1}$). This protocol has been chosen based on its performance in terms of speed and also by the fact that local manipulation of the system's parameters makes it experimentally more feasible. As was the case for the topological chain, the initial state is localized at the first site and corresponds to the system's zero-energy eigenstate (see Fig. \ref{fig:1b} (a)). Here, $J_{1}= 0$ while $J_{i}=J$ $\forall i \neq 1$. During the dynamical evolution, due to the odd-size of the chain, the zero-energy eigenstate is always present. Therefore, gradually switching on $J_{1}$ while $J_{N-1}$ is decreased, results at time $t^{*}$, to the transfer of the excitation at the other end of the chain (see Fig. \ref{fig:1b} (c)). An important difference between this protocol and the one described in the previous section, is that in this case, there is no point in the parameter space where all the couplings acquire the same value. 

\FloatBarrier

\section{Crucial characteristics of the driving function} \label{crucial}

\begin{figure*}
\center
\includegraphics[width=\textwidth]{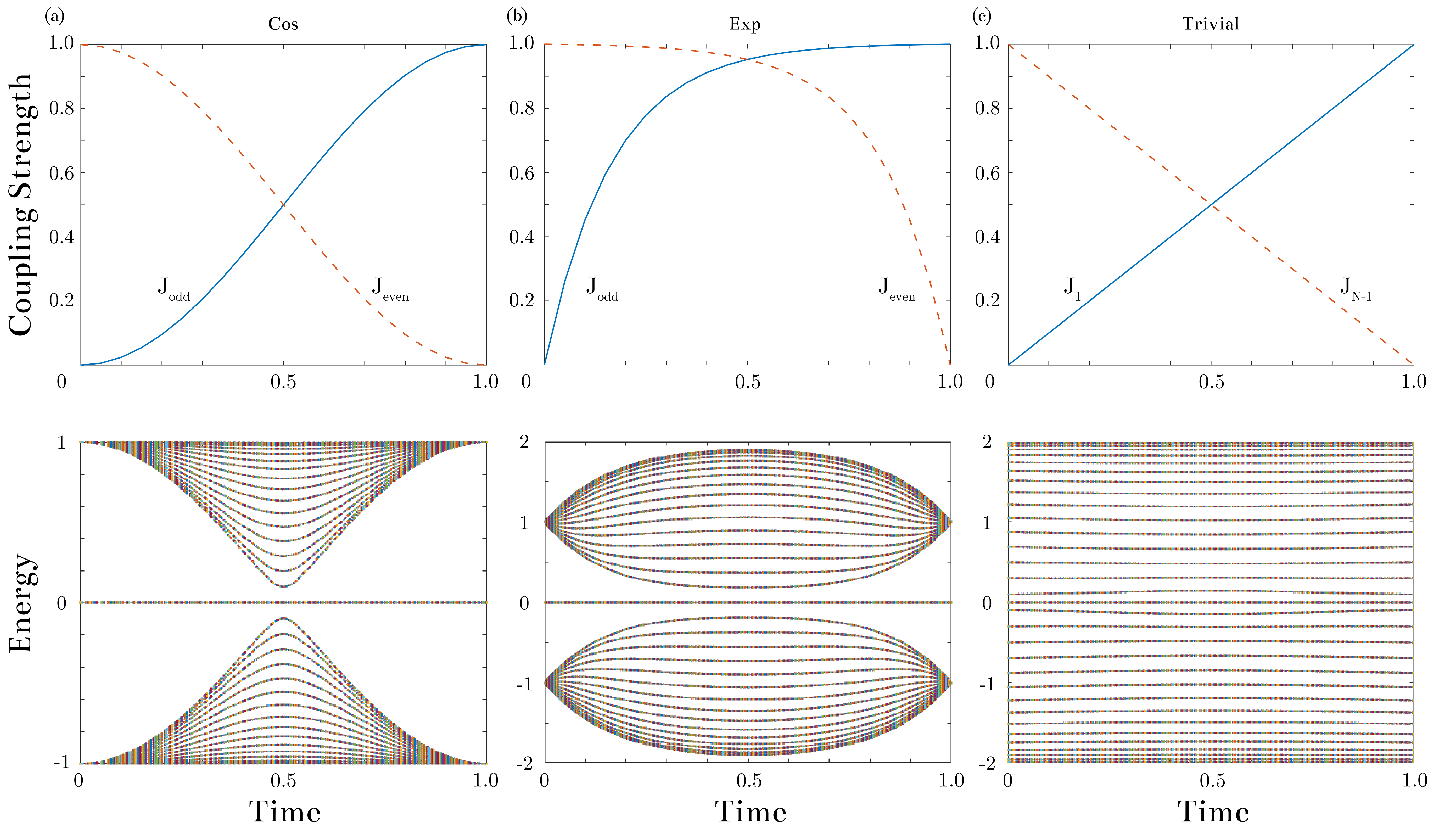}\caption{\label{fig:1} The chain consists of $N=31$ sites. For each protocol (a) cosine (b) exponential and (c) trivial, on the top panel we plot the driving function as a function of time. While on the bottom panel, we depict the corresponding instantaneous energy spectrum as a function of time. In all plots, we have taken the transfer time to be $t^{*}=1$.}
\centering
\end{figure*}

Before presenting our numerical results, we will develop an intuitive and solid line of arguments that dictate which are the crucial considerations that have to be taken into account when driving the state transfer in an odd-sized SSH chain. In all protocols we will consider the system is prepared in the zero-energy eigenstate that is localized on the first site of the chain. As the system evolves, a zero-energy state is always present due to the odd size of the chain. Therefore, the adiabatic approximation ensures, that if the system is driven sufficiently slow during the transfer process, we can remain in the zero-energy eigenstate without exciting bulk modes. What we propose in this paper is to suitably adjust the driving function in order to reach high fidelity values for small transfer times. Our approach does not rely on methods like the adiabatic passage or shortcuts to adiabaticity, where specifically engineered terms are introduced in the Hamiltonian that can induce counter-processes able to suppress the bulk excitations. In other words, we confine ourselves to drive the parameters of nearest-neighbor coupling. To introduce counter-adiabatic terms one should include next-to-nearest neighbor interactions like in \cite{d2020fast}.

When driving the chain, we focus on two important quantities: The energy difference between the zero-mode and the rest of the states, and the derivative of the Hamiltonian matrix that is directly related to the slope of the driving function. These two quantities after all appear in the definition of the adiabatic invariant which is defined as follows: Assuming that $\ket{n,t}$ is the instantaneous eigenmode corresponding to the zero-energy state, in order to be close to the adiabatic limit the following sum has to be sufficiently small
\begin{equation}\label{eq:3}
\sum_{m \neq n} \frac{\bra{m,t}\dot{\hat{H}}\ket{n,t}}{E_{m}(t)-E_{n}(t)} \ll 1,
\end{equation}
where $E_{m}(t)$ the instantaneous eigenenergy of the $m^{th}$ mode and $\dot{\hat{H}}$ the time-derivative of the Hamiltonian. Equation \ref{eq:3} holds when no degeneracies appear in the spectrum and the energies $\abs{E_{m}(t)-E_{n}(t)}>\epsilon_{0}$ are separated by a small $\epsilon_{0}$ $\forall t$.
In the QST protocol employing the odd-sized SSH chain, initially (as $J_{odd}=0$) the energy gap separating the zero-mode from the excited states takes its maximum value. As $J_{odd}$ is switched-on and $J_{even}$ decreases, the energy gap approaches its minimum, occurring at $J_{odd}=J_{even}$ and at the transfer time $t^{*}$ regains its maximum value (e.g. see Fig. \ref{fig:1} (a) and (b) bottom panel). Our logic when dealing with the aforementioned dynamical evolution is simple and can be summarized into two considerations. One thing that we can do is to force the driving function to equate $J_{odd}$ and $J_{even}$ at values close to $J_{max}$, which is the maximum value the couplings can acquire during the transfer. This will result to the maximization of the minimum energy gap. The minimum energy gap can be used to specify a characteristic time scale. When the transfer time is sufficiently large compared to this time scale, we can safely assume that we are close to the adiabatic following of the zero-energy state. A driving function that has this characteristic has also been used in \cite{petrosyan2010state}, where the dimerized chain they consider is equivalent to the SSH chain. The other crucial consideration, is to adjust the driving in such a way that, initially, when the energy gap is bigger, we "strongly" drive the system (i.e. steeper slope of the driving function) and when we are close to the minimum value of the energy gap the driving becomes more "gentle" (i.e. smaller slope). Note however here, that strongly driving the system may induce non-adiabatic transitions between the zero-energy state and the excited states, that in general reduce the efficiency of the transfer.

Our aim is to balance the interplay between the two aforementioned considerations in order to increase the speed of the transfer protocol. An intuitive driving function that we propose and claim to encapsulate this behavior is the exponential. In the next sub-sections it will become clear that the proper implementation of the above leads to faster transfer process while at the same time the robustness of the protocol is maintained.

\FloatBarrier

\subsection{Speed of the transfer} \label{speed}

Now let us examine in more detail the QST protocols that we briefly described in the previous section and provide the numerical evidence supporting our claims. We will examine chains of moderate length $N=31$ sites. In the protocol we propose the couplings are driven by an exponential function (see Fig. \ref{fig:1} (b) top panel), where $J_{odd}= (1-e^{-\alpha t/t^{*}})/(1-e^{-\alpha})$ and $J_{even}=(1- e^{-\alpha (t^{*}-t))/t^{*}})/(1-e^{-\alpha})$, while $\alpha=6.0$ is a free parameter that has been fine-tuned to increase the efficiency in terms of speed. We will get back to the role of this free parameter at the end of the current section. The exponential protocol will be compared with a protocol proposed in \cite{mei2018robust}, where $J_{odd}= b(1-\cos{(\pi t/t^{*})})$ and $J_{even}= b(1+\cos{(\pi t/t^{*})})$ and $b=0.5$ (see Fig. \ref{fig:1} (a) top panel). On the other hand, for the trivial protocol the driving function has the following linear form: $J_{1}=\frac{t}{t^{*}}$, $J_{N-1}=1-\frac{t}{t^{*}}$ and $J_{i}=J_{max}=1$,$\forall i \neq 1,N-1$ (see Fig. \ref{fig:1} (c) top panel). 

In Fig. \ref{fig:1}, we plot for each protocol on the top panel the driving function for the couplings and on the bottom how the instantaneous eigenspectrum evolves over time. Comparing the two protocols that employ the topological chain (see Fig. \ref{fig:1} (a) and (b)), we can immediately notice their qualitative differences. The cosine function initially for large values of the energy gap, drives the system slowly, meaning the numerator of Eq. \ref{eq:3} is smaller as compared to the exponential. However, it approaches the minimum value of the energy gap with greater slope, while the exponential slows down and drives the system more smoothly in this region. Last but not least, the minimum value of the energy gap is analytically obtained by plugging into Eq. \ref{eq:en}, the instantaneous values of the couplings at $t=t^{*}/2$. For the cosine we get $g_{min}^{cos}=0.09$, while for the exponential $g_{min}^{exp}=0.18$. As it was already mentioned, this is because the exponential equates $J_{even}$ and $J_{odd}$ at higher values.

In the trivial protocol on the other hand (Fig. \ref{fig:1} (c)), the evolution is completely different. Namely, the instantaneous energy gap separating the zero-energy mode with the rest of the modes starts from its minimum value ($g_{min}^{triv}=0.1$), slowly increases reaching its maximum ($g_{max}^{triv}=0.14$) at the middle of the time evolution and then returns to its initial value.
\begin{figure}[h]
\center
\includegraphics[width=0.4\textwidth]{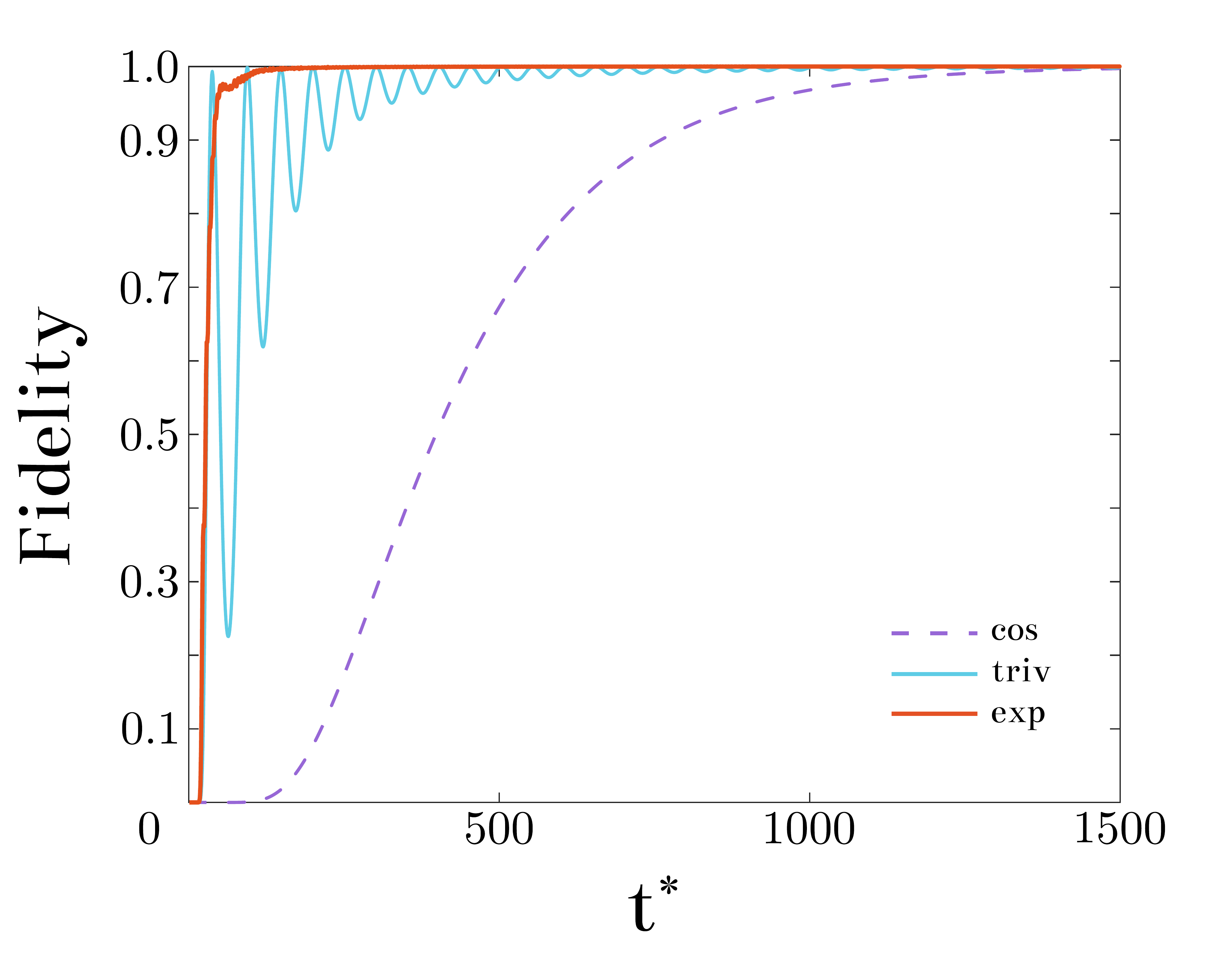}\caption{\label{fig:2} The chain consists of $N=31$ sites. Fidelity as a function of the transfer time for all protocols.}
\centering
\end{figure}

Now that the qualitative differences between the protocols have become apparent let us proceed and examine some quantitative results. In Fig. \ref{fig:2}, for each protocol, we plot the fidelity $\mathcal{F}$ (Eq. \ref{eq:2}) as a function of the transfer time $t^{*}$. To make a comparison in terms of the speed of the transfer we have to set a lower bound in fidelity. In particular, we will consider the time after which the fidelity is stabilized above $0.9$. In this case, the exponential protocol is clearly faster than the cosine protocol, since this occurs for $t^{*}\geq 42$ as compared to the cosine where this happens for $t^{*}\geq 761$. The trivial protocol on the other hand, even though it reaches $\mathcal{F}=0.9$ for $t^{*}=35$, appears to have a strongly oscillatory behavior that prevents its stabilization above $\mathcal{F}=0.9$ till $t^{*} \geq 231$. 

\begin{figure}[h]
\center
\includegraphics[width=0.4\textwidth]{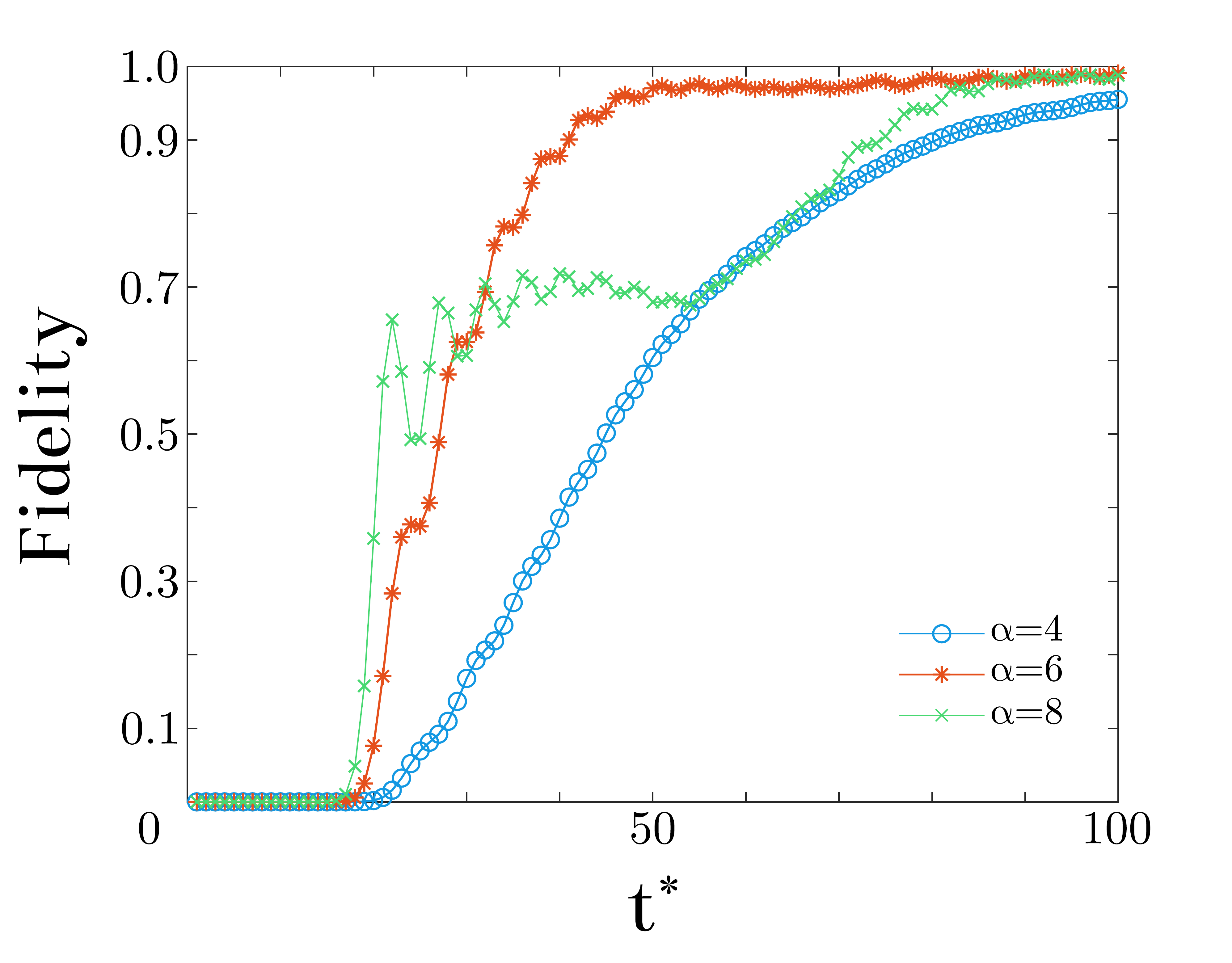}\caption{\label{fig:5} Fidelity as a function of the transfer time for the exponential driving and for different values of the $\alpha$ parameter.}
\centering
\end{figure}

For all profiles, in the limit of $t^{*} \to \infty$ the fidelity approaches unity and the excitation is perfectly transferred along the chain. This makes up the adiabatic limit where, during the dynamical evolution, we "follow" the zero-energy state, without exciting other bulk eigenmodes. The oscillations that appear in the fidelity plot of the trivial protocol are in general unwelcome in QST protocols since they demand great precision when tuning the transfer time \cite{kay2010perfect}. Moreover, they signify that resonant processes are the underlying mechanism responsible for achieving high values of fidelity in such small transfer times. Taking a closer look at the fidelity plot of the exponential protocol (Fig. \ref{fig:5} for $\alpha=6$), we can notice that small oscillations are also present here, i.e. the fidelity curve does not increase smoothly. This indicates that resonant processes are at work also in this case. Obtaining a suitable basis where these processes that occur during the dynamical evolution can be rigorously tracked down, remains a highly non-trivial task \cite{lim1991superadiabatic,dykhne1960quantum}. Nevertheless, as we will now show, the resonant processes can be properly handled to increase the efficiency of the transfer process. 

When we introduced the exponential driving function, we mentioned that the $\alpha$ parameter is fine-tuned ($\alpha=6.0$). Smaller values of the $a$ parameter lead to a less steep slope of the driving function and a smaller value of $g_{min}^{exp}$ (i.e. $J_{even}$, $J_{odd}$ equate at a smaller value). In this case, the resonant processes are suppressed and the fidelity smoothens out (see Fig. \ref{fig:5} $\alpha=4$). Consequently, the protocol's speed is reduced since high fidelity values are obtained for larger transfer times. On the contrary, increasing $\alpha$ above the fine-tuned value results to a stronger slope of the driving function and a greater value of $g_{min}^{exp}$. Therefore, the resonant processes take over for small transfer times and strong oscillations appear at the fidelity plot (see Fig. \ref{fig:5} $\alpha=8$). This once again reduces the speed of the protocol. Thus, the value of this fine-tuned parameter $\alpha=6$ is a trade-off, since it signifies the point up to which we strongly drive the system such that the speed is increased, but gently enough to avoid resonant effects. 

Let us also note here that we have performed optimizations in terms of the CRAB (chopped random basis) algorithm \cite{doria2011optimal} where the guess function was assumed to be the exponential, or the cosine one. The CRAB algorithm gives corrections in terms of a chopped polynomial or Fourier expansion of the correction function. The correction of CRAB to the exponential function was very small, with subtle corrections not changing its basic logic and behavior. On the other hand the cosine function as a guess function in CRAB had a strong correction in the direction of getting closer to the exponential profile. Those calculations indicate that our protocol is close to the optimal one, within the constrains taken in this work.

\FloatBarrier 

\subsection{Disorder Analysis}\label{Disorder}
In this section, we will consider static disorder both on the couplings and on the magnetic field and study its effect on fidelity. Based on the matrix representation of the Hamiltonian the disorder on the couplings is commonly addressed to as off-diagonal disorder, while the disorder on the magnetic field as diagonal disorder. Static disorder can be attributed to manufacturing errors that arise during the experimental implementation. The way each disorder realization is imposed on the system's parameters is the following:
\begin{equation} \label{eq:4}
J_{i}(t) \to J_{i}(t)(1 + \delta J_{i}) \quad B_{i}(t) \to B_{i}(t)+ \delta B_{i}
\end{equation}
$\delta J_{i}$ and $ \delta B_{i}$ acquire random real values uniformly distributed in the interval $(-d_{s},d_{s})$, while $d_{s}$ corresponds to the disorder strength. When we consider static disorder, for each realization a random profile of perturbations is imposed on the parameters and remains fixed during the time evolution.
\begin{figure*}
\center
\includegraphics[width=\textwidth]{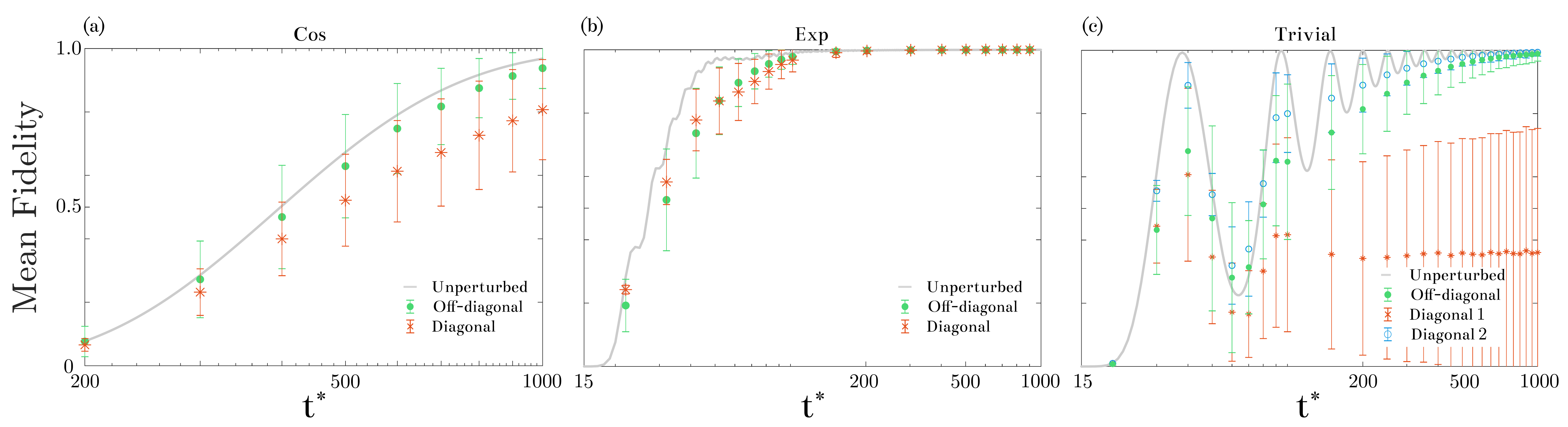}\caption{\label{fig:3} For each protocol, (a) cosine, (b) exponential and (c) trivial, we show the impact of diagonal and off-diagonal disorder of strength $d_{s}=0.2$ (units of $J_{max}$). Each point corresponds to the mean value of fidelity averaged over $10000$ disorder realizations given as a function of the transfer time, while the error bars correspond to the standard deviation of the sample. In order to compare we have also included the unperturbed curve. The transfer time axis is displayed in logarithmic scale. We also note, that the limits of the $t^{*}$-axis for the cosine differ from the other two.}
\centering
\end{figure*}

In Fig. \ref{fig:3} for each protocol, we plot the mean fidelity as a function of $\log{t^{*}}$ for diagonal and off-diagonal disorder of moderate strength $d_{s}=0.2$. What we can immediately notice, is that in almost all cases (there is one exception that will be discussed later on) the effect of disorder does not ruin completely the transfer process. Instead, the main effect is that in the presence of disorder (diagonal or off-diagonal), the transfer time $t^{*}$ needed to reach high values of fidelity is increased. 

Let us turn our attention to the protocols employing the SSH chain. The zero-energy mode of the underlying static chain is known to be robust against perturbations that respect chiral symmetry \cite{asboth2016short}. Off-diagonal (chiral) disorder may change the mode's wavefunction however its energy remains pinned down to zero. On the contrary, diagonal disorder breaks chiral symmetry and the energy of the mode is shifted. For the time-driven chain a difference between chiral and non-chiral disorder becomes apparent in the case of the adiabatic cosine protocol (see Fig. \ref{fig:3} (a)). As it was expected from the static case, the fidelity reduces more in the presence of non-chiral disorder \cite{mei2018robust,lang2017topological}. The exponential protocol however, seems indifferent to whether the disorder is chiral or not (see Fig. \ref{fig:3} (b)). The reason behind this lies in the higher speed of the exponential protocol. Since the effect of disorder strongly manifests in large time scales, the adiabatic cosine protocol is far more sensitive compared to the exponential. This argument has been recently used to justify the resilience of the counter adiabatic protocol in \cite{d2020fast}. 

When we examine the effect of the disorder on the couplings for the topologically trivial protocol (see Fig. \ref{fig:3} (c)) we observe that the oscillatory behavior of fidelity for small transfer times is suppressed (i.e. less oscillations and the mean fidelity is significantly degraded). Thus, we can deduce that the resonant processes are not so robust to the static off-diagonal disorder.

On the other hand, when considering the effect of the disorder on the magnetic field we distinguish two cases. One where the disorder is imposed on all sites of the chain and another where the first and last sites are exempted (i.e. $\delta B_{1}=\delta B_{N}=0$). In the latter case, the protocol proves to be even more robust than the off-diagonal case (see Fig. \ref{fig:3} (c) diagonal 2). However, in the former, the effect of disorder is severe and the transfer process is completely destroyed (see Fig. \ref{fig:3} (c) diagonal 1). The diagonal disorder on the edges greatly affects the transfer process since it induces an energy difference between the initial and the final state. The system's initial state is localized on the first site with energy equal to $\delta B_{1}$, while the final state is localized on site $N$ with energy $\delta B_{N}$. Combined with the fact that the energy gap separating them from the rest of the excited states takes its minimum value (which is smaller than the strength of the disorder) during the beginning and the end of the transfer process, explains the high impact of the diagonal disorder on the edges. In conclusion, the topological protection coming from the energy gap, clearly favors the topological channels, which are indifferent to whether the diagonal disorder is imposed on the edge sites. 

To sum up, the exponential protocol is quite robust to both on and off-diagonal disorder and clearly outperforms the two other protocols. As opposed to the adiabatic cosine protocol, it is indifferent to whether the disorder is chiral or not and compared with the trivial chain we can deduce that no significant (in terms of affecting the fidelity) resonant processes susceptible to static noise are at work. Finally, the presence of a wide energy gap in the underlying static SSH chain clearly favors the topological quantum channel when the diagonal disorder is imposed on the edge sites of the chain. 

\FloatBarrier

\section{Conclusions} \label{Conclusions}
In this work we have numerically investigated a time-dependent protocol that employs a topological quantum chain to act as a quantum channel for transferring single-site excitations. We propose an exponential driving function that increases the efficiency of the transfer in terms of speed. To sustain our claim, we make a comparison with two other QST protocols. The crucial characteristics of the exponential function are the fact that it suitably adapts the slope of the driving function based on the value of the instantaneous energy gap, while at the same time ensures that the minimum value of the energy gap $g_{min}$ is as big as possible. The resonant processes are fine-tuned, leading to a speed increasal. Employing the CRAB optimization algorithm, we get strong indications that the proposed time-driving function is close to optimal. In addition, we study the effect of diagonal and off-diagonal static noise highlighting the fact that even though the speed of the protocol is increased its robustness is maintained. The difference in terms of speed with the cosine and the trivial protocol, emphasizes the power of our treatment and identifies the considerations that have to be taken into account when driving a topological quantum chain. The developed scheme adds up to the ongoing effort of constructing discrete networks that can efficiently transfer and manipulate quantum states. It also makes a substantial contribution to speeding-up adiabatic protocols (with topological characteristics or not) since it indicates a conceptual way of designing the control schemes depending on the instantaneous eigenspectrum characteristics.

\section*{Acknowledgments}
G. T. and I. B. acknowledge funding  by the project CS.MICRO funded under the program Etoiles Montantes of the Region Pays de la Loire. N. E. P. gratefully acknowledges financial support from the Hellenic Foundation for Research and Innovation (HFRI) and the General Secretariat for Research and Technology (GSRT), under the HFRI PhD Fellowship Grant No. 868.

\end{document}